\documentclass[11pt, oneside]{article}  	
\usepackage[left=2cm,top=3cm,right=2cm,bottom=4cm,head=1cm]{geometry}     
\usepackage[utf8]{inputenc}
\usepackage{graphicx}				
\usepackage{amsmath}
\usepackage{multicol}
\usepackage{amssymb}
\usepackage{titling}
\usepackage{hyperref}
\usepackage{authblk}

\numberwithin{equation}{section}
\usepackage{wrapfig}
\usepackage[justification=centering]{caption}
\newcommand{\drv}{\ensuremath{\textup{d}}} 

\title{\textbf{Forces Between Kinks in $\boldsymbol \phi^\textbf{8}$ Theory}}
\date{\small October 2019}
\author[1]{Peru d'Ornellas}

\affil[1]{\small Blackett Laboratory, Imperial College London, London SW7 2AZ, United Kingdom}

\begin{document}

\maketitle

\begin{abstract}
    We investigate the dynamics of the kinks that emerge in a one-dimensional scalar field theory with an octic potential containing a quartic minimum and two quadratic minima. We show analytically that kink-antikink and kink-kink pairs interact with a force that scales with the fourth power of the inter-kink distance, and calculate its strength. This is done using two different techniques. The first employs a collective coordinate method to approximately solve the equation of motion for the profile of an accelerating kink. The second is based on modifying the potential to one that is able to support static solutions containing multiple kinks. We show that the two methods give consistent results. All calculations are supported by numerical work that confirms the validity of our results.
\end{abstract}

\section{Introduction}
Kinks are the simplest example of a topological soliton, appearing in one-dimensional field theories with a single scalar field and a potential with multiple global minima \cite{manton_topological_2004,rajaraman_solitons_2005}. They appear as static solutions to the equation of motion that interpolate between the different minima and are an example of a topologically protected state, where to remove them requires one to change some topological characteristic of the field. Thus, it is generally not possible for the kink to be removed or added as a field evolves smoothly under the equation of motion. Furthermore, they appear with finite spatial extent and well-defined position and mass, so it is possible to view them as particle-like objects, capable of moving through space and interacting with other kinks as a particle might.\par
Kinks, as well as higher-dimensional solitons such as vortices and skyrmions, have found applications in an enormous range of physical systems. They have appeared in condensed matter \cite{bishop_solitons_1979,chaikin_walls_1995}, where kinks can be used as a one-dimensional model for the interface separating different phases of matter, in cosmology \cite{vilenkin_cosmic_2001} and in high-energy physics \cite{manton_topological_2004,weinberg_classical_2012} where they have been used as models for elementary particles.\par
In this work we will be focussing on the kinks that appear in a $\phi^8$ theory \cite{manton_forces_2019,christov_long-range_2019}, where the Lagrangian is given by 
\begin{equation}
    \mathcal{L} = \frac{1}{2} \partial_\mu \phi \partial^\mu \phi  - \frac{1}{2} \left ( 1- \phi^2 \right )^2 \phi^4.
\end{equation}
Unlike the kinks in other well-studied models such as the $\phi^4$ model \cite{lohe_soliton_1979,kevrekidis_dynamical_2019} or the Sine-Gordon model \cite{manton_topological_2004,rajaraman_solitons_2005}, this field theory contains kinks with large spatial extent, capable of interacting over very long distances. One of the primary questions here has been to calculate the strength of interaction between well-separated kinks. This has been investigated both analytically \cite{manton_forces_2019,gonzalez_kinks_1989} as well as computationally \cite{christov_long-range_2019,belendryasova_scattering_2019}. In particular, obtaining accurate parameters for the interaction strength through numerical work has proved much more challenging than in the case of short-range kinks. This is because $\phi^8$ kinks have a strong, asymmetric sensitivity to the presence of radiation. This means that one must take extreme care when picking the initial configuration of the field to directly simulate the dynamics of a kink. Incorrectly initialised fields will result in unwanted radiation being produced that substantially disrupts the dynamics.\par
In this article we present two different methods for calculating the interaction strength between kinks in our theory, supporting both with numerical work. The first closely follows \cite{manton_forces_2019}, supporting the analytical work presented there with numerical calculations that verify its validity. Importantly, this numerical work avoids many of the pitfalls associated with directly simulating the dynamics of the field. We follow this with analysis that extends the ideas from \cite{gonzalez_solitary_1987,gonzalez_kinks_1989} to accurately predict the interaction strength. This work is supported by numerical calculations that confirm that it is sound. Furthermore, we show analytically that the two unrelated methods are consistent with one another.

\section{Setting the Scene}\label{section:set_scene}
We begin the discussion by defining the one-dimensional field theory in question and deriving some of the fundamental properties of the solitons that emerge in this theory, namely the shape of the kink, including the asymptotic behaviour of its tails and its mass.\par
The Lagrangian density for a scalar field $\phi(x)$ with potential $V(\phi)$ is given by
\begin{equation}\label{lagrangian}
    \mathcal{L} = \frac{1}{2} \partial_\mu \phi \partial^\mu \phi  - V(\phi).
\end{equation} 
We are working in Minkowski space, and will use the metric $\eta_{\mu\nu} = \textup{diag}(+1,-1)$, thus the Lagrangian may be rewritten as
\begin{equation}
    \mathcal{L} = \frac{1}{2}\dot{\phi}^2 -  \frac{1}{2}{\phi'}^2   - V(\phi).
\end{equation}
Here dot denotes derivative with respect to time and prime represents spatial derivative.\par
The Euler-Lagrange equation is the equation of motion (EOM) for the field,
\begin{equation}\label{EOM}
    \ddot{\phi} - {\phi}'' + \frac{\drv V}{\drv \phi} = 0.
\end{equation}
The Lagrangian does not depend explicitly on the spacetime coordinates. Thus we may invoke Noether's theorem to find the conserved energy-momentum tensor,
\begin{align}
    T_{\mu \nu} &= \frac{\partial \mathcal{L}}{\partial \left (\partial^\mu \mathcal{\phi}\right )} \partial_\nu\phi - \eta_{\mu\nu} \mathcal{L}  \nonumber \\
    &=  \partial_\mu\phi \partial_\nu\phi - \eta_{\mu\nu} \mathcal{L}  
\end{align}
with $\partial_{\mu}{T^{\mu}}_{\nu} = 0$.\par
In particular, the energy density $\mathcal{E}$ and momentum density $\mathcal{P}$ may be read directly from components of the tensor according to
 \begin{align}
    T_{00} &= \frac{1}{2}\dot{\phi}^2 +  \frac{1}{2}{\phi'}^2   +V(\phi) = \mathcal{E} \label{energy}\\
    T_{01} &= \dot{\phi} \phi'  = -\mathcal{P}. \label{momentum}
\end{align}\par
\begin{figure}[t]
    \vspace{-10pt}
    \centering
    \includegraphics[width=0.45\textwidth]{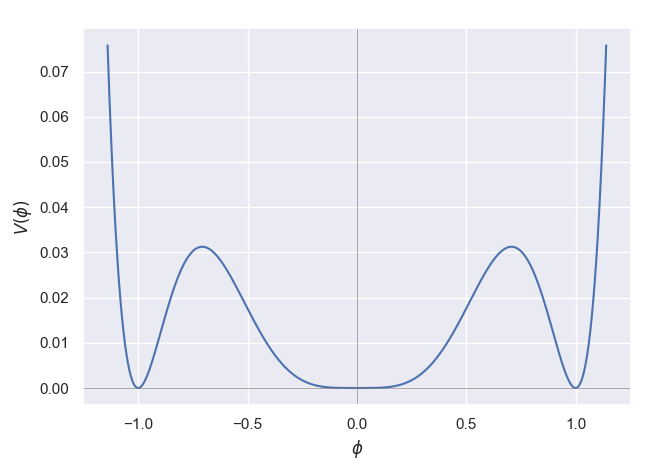}
    \captionof{figure}{Potential in $\phi^8$ theory} \label{phi8pot}
\end{figure}
In this work we focus on kinks that emerge in the $\phi^8$ potential
\begin{align}
    V(\phi) = \frac{1}{2} \left ( 1 - \phi^2\right)^2 \phi^4,
\end{align}
plotted in fig.~\ref{phi8pot}. The potential has three minima, a pair of quadratic minima at $\phi = \pm1$ and a quartic minimum at $\phi= 0$. Thus it is able to support four different types of kink. We will label kinks as interpolating from $\phi = 0 \rightarrow 1$, antikinks from $\phi = 1 \rightarrow 0$, mirror kinks from $\phi = -1 \rightarrow 0$ and anti-mirror kinks from $\phi = 0\rightarrow -1$.\par
Our first task is to derive the form of the kink appearing in this theory, using the Bogomolny trick \cite{bogomolny_stability_1976}. The kink is a static solution to the EOM (eqn.~\ref{EOM}). This means that the time derivative may be discarded, resulting in the following nonlinear ODE:
\begin{equation} \label{kink_equation}
    {\phi}'' - \frac{\drv V}{\drv \phi} = 0.
\end{equation}
The potential may be re-expressed in terms of a superpotential $W$ according to
\begin{equation}
    V = \frac{1}{2} \left ( \frac{\textup{d}W}{\textup{d}\phi}\right )^2
\end{equation}
with
\begin{equation}\label{dw}
    \frac{\textup{d}W}{\textup{d}\phi} = \left ( 1 - \phi^2 \right ) \phi^2.
\end{equation}
Thus we have
\begin{align}
    W = \frac{1}{3}\phi^3 - \frac{1}{5}\phi^5 +c.
\end{align}
For the sake of later convenience we set $c$ to be $-\frac{2}{15}$, so that $W[1]=0$. We expect the kink solution to have $\phi (-\infty )= 0$ and $\phi (\infty )= 1$. Furthermore, the solution should be a minimum of the energy. Using the expression for energy density obtained in eqn.~\ref{energy}, we can write the total energy of the field as
\begin{equation}\label{EW}
    E = \frac{1}{2} \int_{-\infty}^{\infty} \left [ \left(\frac{\textup{d}\phi}{\textup{d}x}\right)^2 + \left(\frac{\textup{d}W}{\textup{d}\phi}\right)^2 \right ]\drv x.
\end{equation}
This can be rearranged by completing the square to give two equivalent expressions, depending on sign,
\begin{equation}\label{energy_bogo}
    E = \frac{1}{2} \int_{-\infty}^{\infty}\left (  \frac{\textup{d}\phi}{\textup{d}x} \mp \frac{\textup{d}W}{\textup{d}\phi} \right )^2 \drv x  \, \pm \left( W\left [  \phi(\infty)\right ]  -  W\left [  \phi(-\infty)\right ]\right).
\end{equation}
$\phi(\pm\infty)$ will always be an element of $\Phi_{\textup{vac}} = \left \{ -1,0,+1\right \}$. From our expression for $W$ it can be shown that $W[-1] = -\frac{4}{15}$, $W[0] = -\frac{2}{15}$ and $W[1] = 0$.\par
\begin{figure}[t]
    \centering
        \includegraphics[width=0.45\textwidth]{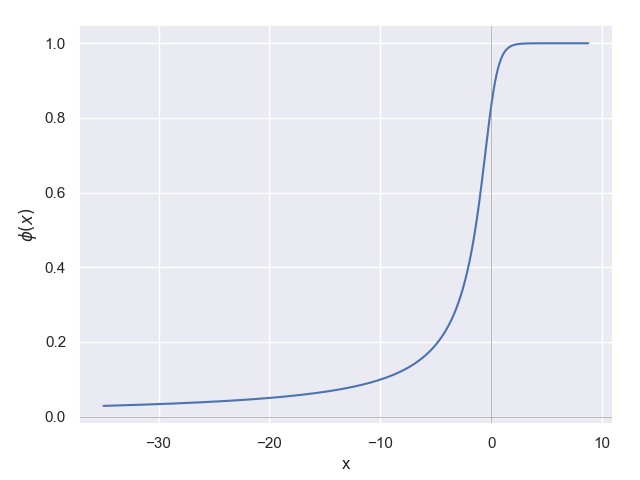} 
    \captionof{figure}{Kink in $\phi^8$ theory} \label{phi8kink}
\end{figure}

Equation \ref{EW} is a sum of two squares, so $E\geqslant 0$. This means that we should pick the sign to ensure that $ \pm \left( W\left [  \phi(\infty)\right ]  -  W\left [  \phi(-\infty)\right ]\right) \geqslant 0$ for our choice of limits. In the case of the kink, $\phi(-\infty) = 0,\, \phi(\infty) = 1 $, therefore $W\left [  \phi(\infty)\right ]  -  W\left [  \phi(-\infty)\right ] = \frac{2}{15}$ and so we pick the expression for the energy with a minus sign inside the integral. The total energy is minimised when the expression in the integral is 0, that is when
\begin{equation}\label{bogo}
    \frac{\textup{d}\phi}{\textup{d}x} - \frac{\textup{d}W}{\textup{d}\phi} = 0.
\end{equation}
This is called the Bogomolny equation. Using eqn.~\ref{dw} it can be rearranged to give
\begin{equation}
    \int \frac{\drv\phi}{\left(1-\phi^2 \right ) \phi^2} = \int \drv x
\end{equation}
which may be separated into partial fractions,
\begin{equation}
    \int\drv\phi \left (\frac{1}{2\left(1+\phi\right )} + \frac{1}{2\left(1-\phi\right )} + \frac{1}{\phi^2} \right) = x -A,
\end{equation}
and integrated to give an implicit expression for the shape of a kink
\begin{equation}\label{long_kink_eq}
    x-A = \frac{1}{2}\log\left (\frac{1+\phi}{1-\phi} \right ) - \frac{1}{\phi}.
\end{equation}
Here the parameter $A$ determines the position of the kink. This expression must be numerically inverted to give the form of the kink, shown in fig.~\ref{phi8kink}, as there is no explicit expression for the inverted kink profile, $\phi(x)$. It is instructive to examine the asymptotic behaviour of the kink for $\phi$ close to 0 and close to 1. \par
On the left hand side of the kink, $\phi = \epsilon$ with $\epsilon$ small and positive. In this case, eqn.~\ref{bogo} reduces to $\frac{\drv \epsilon}{\drv x} = \epsilon^2$, which can be solved to give 
\begin{equation}
    \phi_\textup{left} = \frac{1}{A-x}.
\end{equation}
Note that $\phi_\textup{left}$ diverges at $x = A$. On the right hand side $\phi = 1- \epsilon$ with $\epsilon$ small and positive and eqn.~\ref{long_kink_eq} can be expanded to first order to give
\begin{equation}
    \frac{1}{2} \log2 - \frac{1}{2} \log \epsilon - 1 = x-A
\end{equation}
which gives the following expression for $\phi_{\textup{right}}$,
\begin{equation}
    \phi_{\textup{right}} = 1- \exp\left [ -2\left ( x - A + 1 - \frac{1}{2}\log2 \right ) \right ].
\end{equation}
We can see that in the case of $\phi^8$ theory, the decay on the left of the kink has an extremely long tail, allowing for long-range interactions.\par
To talk about the position of a kink, we will need a convention defining the center of a kink. The kinks in $\phi^8$ theory have large spatial extent, as well as not being symmetric in $x$. Thus there is no obvious definition for the centre of a kink and we must make a somewhat arbitrary choice. In this work we define the center of the kink as the point that crosses the maximum of the potential energy. This is the point where the energy density of the kink is highest and occurs at $V(\phi) = V_{\textup{max}}$,
\begin{align}\label{eqn:kink_center}
    \phi_{\textup{center}} = \frac{1}{\sqrt{2}}.
\end{align}
Throughout this work, we will only be concerned with the interactions between kinks in the long-range limit, so the specific choice of kink center will be of little importance.
\begin{figure}[t]
    \vspace{-10pt}
    \centering
        \includegraphics[width=0.45\textwidth]{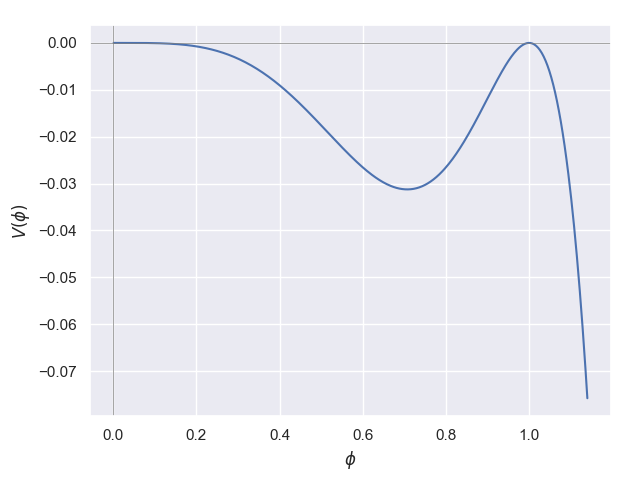}
    \captionof{figure}{Inverted potential felt by the particle in the mechanical reinterpretation} \label{mech}
\end{figure} \par
Here it is worth introducing the mechanical reinterpretation of the kink solution, since it will be useful later on. Equation \ref{kink_equation} may be interpreted as the equation of motion for a `particle' with `position' $\phi$ moving in a potential $-V(\phi)$. To produce a kink, the particle starts at $\phi = 0$, the maximum of the potential, with infinitesimal positive velocity. It moves quickly through the trough and then slows down to asymptotically approach the next maximum at $\phi = 1$. Energy is conserved in this system (there is no drag term), so the particle will come to rest at the next maximum. As the particle moves, it sweeps out the profile of the kink.\par
Finally we discuss the mass of the kink. From eqn.~\ref{energy_bogo} we can see that, once the Bogomolny equation is satisfied, the total energy of a kink is $\frac{2}{15}$. Thus, since the theory is Lorentz invariant, the kink may be interpreted as a particle-like object with mass $\frac{2}{15}$. Furthermore, since the Bogomolny equation is satisfied locally at every point in space, the total energy of the field between two points $x_1$ and $x_2$ is $\left | W\left [ \phi(x_1)\right ] - W\left [ \phi(x_2)\right ]\right |$.

\section{Forces between Accelerating Kinks}
We now turn our attention to modelling the force between a well-separated pair of kinks. In \textsection \ref{kink-antikink-manton} we look at the attractive force between a kink and antikink. Then in \textsection \ref{sec:kink-mirrorkink-manton} we calculate the force of repulsion between a kink and mirror kink. In all cases the kinks are ordered such that the interaction is due to the overlapping of their long-range $\frac{1}{x}$ tails. The short-range tail on the other side has exponential asymptotics, so interactions where these tails overlap will be largely identical to those in $\phi^4$ theory. This theory is well-understood and so we do not study it here \cite{manton_topological_2004}. The force is initially calculated analytically by approximately deriving a profile for the field containing a pair of interacting kinks. By examining the momentum density of the field, it is possible to extract the force acting on the pair. The analytical discussion closely follows \cite{manton_forces_2019}. In each case, we support the analysis with numerical results that verify that the method is sound and the approximations made are valid.\par

\subsection{Kinks and Antikinks}\label{kink-antikink-manton}
\begin{figure}[t]
    \vspace{-10pt}
    \centering
    \includegraphics[width=0.45\textwidth]{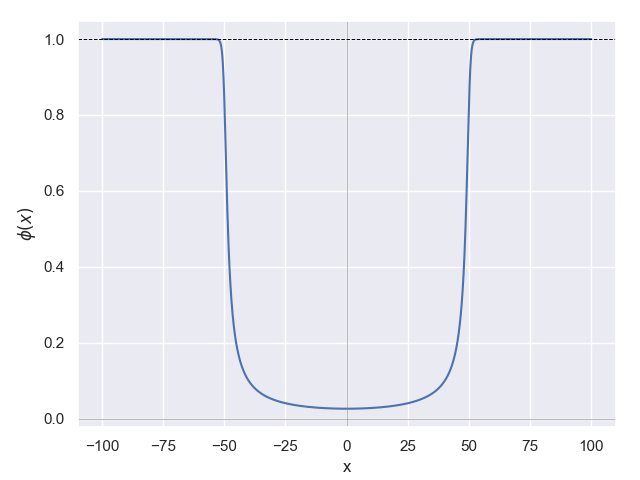}
    \captionof{figure}{A possible state containing a kink and anti-kink} \label{kink_kink_state}
    \vspace{-10pt}
\end{figure} 
To calculate the force between an interacting kink and antikink pair, we must make a guess at an appropriate field that contains these two objects. The system we are looking at is no longer static -- the kinks are accelerating towards one another -- so it will be necessary to account for the distortion of the kink profile caused by this acceleration. Let us start with a field in which the antikink is at position $-A(t)$, the kink is at $A(t)$, and the field is initially at rest.\par
The kink and antikink have identical form, but reflected in $x$, therefore it is reasonable to propose a configuration that is symmetric, as shown in fig.~\ref{kink_kink_state}. Here the antikink is positioned at $x = -50$ and the kink is at $x= 50$. Note that the field between the kinks does not touch the $x$-axis. The field is distorted from the static kink solution, with $\phi' = 0$ at $x=0$. It is possible to construct such a solution by finding the profile for a single accelerating kink and gluing together the tails of the accelerating kink and its reflection.\par
Before calculating the shape of this field, we first examine how the force acting on a kink may be determined from the field configuration. From eqn.~\ref{momentum} it can be seen that the momentum density is given by $\mathcal{P} = -\dot{\phi}\phi'$. Therefore, the rate of change of the total momentum in a region from $x_a$ to $x_b$ is given by
\begin{align}
    \frac{\drv P}{\drv t} &= -\frac{\drv}{\drv t} \int_{x_a}^{x_b} \dot{\phi}\phi' \drv x \\
    &= - \int_{x_a}^{x_b} \left ( \ddot{\phi}\phi' + \dot{\phi}\dot{\phi'} \right )\drv x.
\end{align}
From the equation of motion, $\ddot{\phi} = \phi'' - \frac{\drv V}{\drv\phi}$, this expression becomes
\begin{equation}
    \frac{\drv P}{\drv t} = - \int_{x_a}^{x_b} \left ( \phi'' \phi' - \frac{\drv V}{\drv\phi} \phi' + \dot{\phi}\dot{\phi'} \right )\drv x
\end{equation}
which may be integrated to give an expression for the force acting on the region between $x_a$ and $x_b$
\begin{equation}\label{force_total}
    F = -\left [ \frac{1}{2} \phi'^2 + \frac{1}{2} \dot{\phi}^2 - V(\phi) \right ]_{x_a}^{x_b}.
\end{equation}
In the case of two interacting kinks, shown in fig.~\ref{kink_kink_state}, the force can be determined by setting $x_a = 0$ and $x_b = \infty$. The field at $ x =\infty$ is in the ground state and does not contribute to the force. We expect the field at $x=0$ to have no spatial first derivative and negligible time derivative, since the field starts from rest and is initially slow moving. Thus we are left with only one contributing term
\begin{equation} \label{force}
    F_\textup{kink} = - V[\phi(0)].
\end{equation}\par
Now we return to determining a form for the field of a single accelerating kink. The first step is to model the kink with an expression of the form
\begin{equation}
    \phi(x,t) = \chi\left( x - A(t)\right).
\end{equation}
This can also be written as $\chi\left( y\right)$ for $y = x-A(t)$. Substituting into the equation of motion gives
\begin{equation}
    \chi '' \dot{A}^2 - \chi'\ddot{A} - \chi '' + \frac{\drv V }{\drv \chi}=0.
\end{equation}
We are starting with a field at rest, so $\dot{A}$ is negligible, thus the term $\chi '' \dot{A}^2$ may be ignored. Furthermore we assume the acceleration is small and constant, allowing us to set $\ddot{A} = -a$,
\begin{equation} \label{acc_kink}
    \chi'' - a\chi' - \frac{\drv V }{\drv \chi}=0.
\end{equation}
In the mechanical interpretation, this may be understood as describing the dynamics of a particle moving in the same potential as fig.~\ref{mech} with the addition of a negative drag term with coefficient $-a$. The effect of $a$ is to add energy to the system as the particle moves under the influence of $V(\chi)$. The solution we are looking for corresponds to the `particle' starting from rest ($\chi' = 0$) at a point $\chi >0$ and gaining just enough energy over its motion to make it to maximum in the potential at $\chi=1$, where it comes to rest. For each value of $a$ there is a position $\chi(0)$ such that a particle starting at that point and moving under drag $-a$ will gain just enough energy to reach $\chi=1$ and go no further, resulting in a valid accelerating kink solution.\par
\begin{figure}
    \centering
    \includegraphics[width=0.9\textwidth]{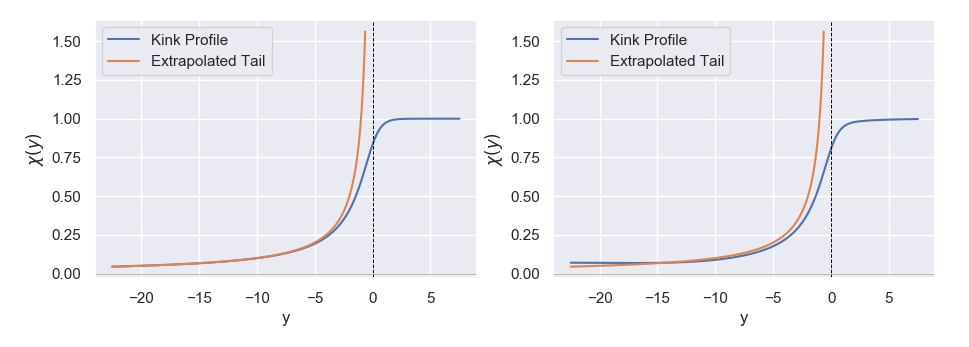}
    \captionof{figure}{Static and accelerating kink with extrapolated tail behaviour. The points at which the extrapolated tails diverge are marked with a dashed line.} \label{tails}
\end{figure}

In what follows, we assume that the majority of the influence of the drag term is exerted on the long-range tail of the kink. This makes sense since, at least in the limit of well-separated kinks, the tail is extremely long. The drag force is weaker in the tail region than it is around the centre of the kink, but not substantially so, and acts over a much larger region. To compare the accelerating kink with the static kink derived in \textsection \ref{section:set_scene} we match the shapes of the long tails of both, in particular, aligning them so that the extrapolation of their $\frac{1}{x}$ tails diverge at the same point. This is shown in fig.~\ref{tails}, where the point at which the tail diverges is marked with a dashed line.\par
We can assume that $\chi$ approximately takes the form of an undeformed kink over its long tail, ensuring that it is a solution of the Bogomolny equation ($\chi' = \frac{\drv W }{\drv \chi}$), with $W$ defined in eqn.~\ref{dw}. Thus, eqn.~\ref{acc_kink} becomes
\begin{equation}\label{VW}
    \chi'' -\frac{\drv}{\drv \chi}\left ( aW + V \right ) =0.
\end{equation}
Now we define a modified potential $\tilde{V} = V +aW$ and integrate eqn.~\ref{VW},
\begin{equation}
    \int \chi'' \chi' \drv x = \int \frac{\drv \tilde{V}}{\drv \chi} \drv \chi
\end{equation}
to get
\begin{equation}
    \chi'^{\,2} = 2 \tilde{V}.
\end{equation}
In the long-tail region, $\chi$ is small, so $V\simeq \frac{1}{2}\chi^4$ and $W\simeq -\frac{2}{15}$, thus we obtain
\begin{equation}\label{chi_prime}
    \chi' = \sqrt{\chi^4 - \frac{4a}{15}}.
\end{equation}
$\chi$ takes its smallest value at $x=0$ where $\frac{\drv\chi}{\drv x}=0$. Therefore for eqn.~\ref{chi_prime} to hold we must have $\chi(-A) = \left( \frac{4a}{15} \right)^{1/4}$ (where we used $y = x-A$). Note that from eqn.~\ref{force} this means that the force acting on the kink is given by
\begin{equation}\label{force_acceleration}
    F = -V\left[\chi(0)\right] = -\frac{2a}{15},
\end{equation}
which is perfectly consistent with Newton's law of motion for a kink of mass $\frac{2}{15}$ and acceleration $-a$. We wish to fix the solution of eqn.~\ref{chi_prime} to ensure that it diverges at $x = A$ ($y = 0$) since the extrapolation of a static kink diverges at $x=A$. We therefore have $\chi(0) = \infty$. Having set our limits, we can integrate eqn.~\ref{chi_prime} to get
\begin{equation}
    \int_{ \left( \frac{4a}{15} \right)^{\frac{1}{4}}}^{\infty}\frac{\drv \chi}{\sqrt{\chi^4 - \frac{4a}{15}}} = A.
\end{equation}
Making the substitution $\chi =\left( \frac{4a}{15}\right)^{\frac{1}{4}}\lambda$ gives
\begin{equation}
    \int_{1}^{\infty}\frac{\drv \lambda}{\sqrt{\lambda^4 - 1}} = \left( \frac{4a}{15}\right)^{\frac{1}{4}} A.
\end{equation}
The left-hand side is a complete elliptic integral of the first kind, and can be evaluated to give \cite{gradshtein_table_2007}
\begin{equation}
    \left( \frac{4a}{15}\right)^{\frac{1}{4}} A =\frac{1}{\sqrt{2}}\frac{\Gamma \left( \frac{1}{4}\right)^2}{4\sqrt{\pi}}.
\end{equation}
Plugging this result into eqn.~\ref{force_acceleration} and evaluating gives
\begin{equation} \label{prediction}
    F =-\frac{1}{8A^4}\left (\frac{\Gamma(\frac{1}{4})^2}{4\sqrt{\pi}}  \right )^4 \simeq -\frac{1.47713...}{A^4}.
\end{equation}

\subsubsection{Numerical Investigation}\label{sec:numerical_manton_kak}
A number of studies have been undertaken to verify the relationship derived above by numerically simulating the dynamics of the field \cite{christov_long-range_2019,belendryasova_scattering_2019,christov_kink-kink_2019}.  We take a different approach to finding the strength of the force acting between a kink and anti-kink, focussing on carefully approximating the profile of an accelerating kink. Our starting point is eqn.~\ref{acc_kink}:
\begin{equation}
\chi'' - a\chi' - \frac{\drv V }{\drv \chi}=0.
\end{equation}
Note that in deriving this differential equation, the only approximation we made was to assume that $\dot{A}$ can be ignored, thus it is exact for finding the force acting on a static kink, before it has started to move under acceleration.\par
\begin{figure}[t]
\vspace{-10pt}
\centering
\includegraphics[width=0.45\textwidth]{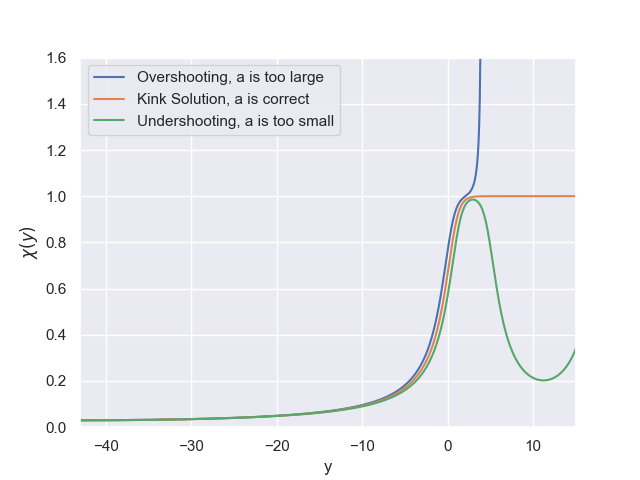}
\captionof{figure}{Plot of the particle's motion for three different values of $a$, with $\chi_0 = 0.003$.} \label{over_under}
\vspace{-10pt}
\end{figure}

We examine the differential equation through the lens of the mechanical interpretation, viewing it as describing a particle that starts at a position $\chi(-A)$ and moves under negative drag term $-a$ in the potential $- \frac{\drv V }{\drv \chi}$. As explained in the previous section, for every value of $\chi(-A)$ there is a unique corresponding value of $a$ such that the motion of the particle will correspond to a valid kink solution. If $a$ is too large, the particle will gain too much energy as it moves through the potential, overshooting the maximum and `falling off' the other side, accelerating towards $\chi \rightarrow +\infty$. If $a$ is too small, the particle will not gain enough energy as it moves through the potential. This means it won't make it up to the maximum of the potential at  $\chi=1$ and will turn back on itself, moving back towards negative $\chi$. Fig. \ref{over_under} shows the motion of the particle for three values of $a$, one with an overshoot, one that is correct and one with an undershoot.\par
This differential equation is easily solved using python's inbuilt ODE solver, \texttt{odeint}. We start with a choice of initial values, $\chi(-A)$, $\chi'(-A) = 0$ and $a$, and solve it to obtain a profile for $\chi(y)$, with $y$ discretised over an array of length 2048. It is straightforward to determine if the value of $a$ was an overshoot or an undershoot, so we can quickly converge to the correct value using a method based on the binary search algorithm. \par
Once we have determined the value of $a$, we automatically get an expression for the shape of an accelerating kink for our chosen $\chi(-A)$ as the solution of our differential equation. Now the only thing left to do is to find the center of the kink, $A$, by looking for the point in space where the kink crosses $\chi(0) = \frac{1}{\sqrt{2}}$. The force acting on the kink is given by eqn.~\ref{force}, $F = -V[\chi(-A)]$. Thus we have arrived at a set of values for the force acting on the kink as a function of the kink separation.\par
\begin{figure}[t]
    \vspace{-10pt}
    \centering
    \includegraphics[width=0.42\textwidth]{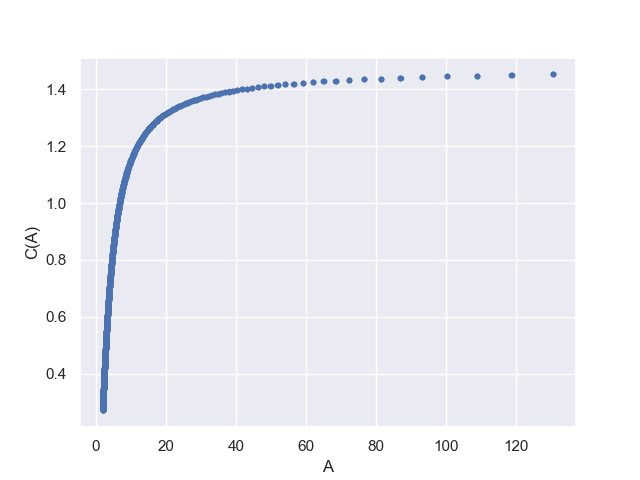}
    \captionof{figure}{Plot showing the relationship between force coefficient $C(A)$ and inter-kink separation $A$ for an interacting kink-antikink pair.} \label{coefficient_scatter}
    \vspace{+10pt}
\end{figure}
This process was performed for a set of 8000 values of $\chi(-A)$ spaced evenly between $\chi(-A) = 0.01$ and $\chi(-A) = 0.5$. Then, using the 8000 pairs of values for force $F$ and distance $A$, the force coefficient
\begin{align}\label{force_coeff}
    C(A) = FA^4
\end{align}
was calculated. Based on the results from the previous section, it is expected that for large separations $C(A)$ should tend towards a constant value (around 1.477...), but will deviate from this for smaller values of $A$. We fit to an expression of the form
\begin{table}[t]
    \begin{center}
    \caption{$c_0$ and residuals for varying $n$}\label{table:kink_antikink_different_fittings}
        \begin{tabular}{|c @{ : } c c c c c c|}
        \hline
        $n$ &$ 0 $ &$ 1 $ &$ 2 $ &$ 3 $ &$ 4 $ &$ 5 $\\
        $c_0$& 0.81562 & 1.38717 & 1.49764 & 1.48974 & 1.47978 & 1.47695\\
        $S$ &0.13 & 0.0026 & 1.6$\times 10^{-05}$ & 8.4$\times 10^{-06} $& 4.1$\times 10^{-07}$ & 2.6$\times 10^{-09}$\\
        \hline
        \end{tabular}
        \begin{tabular}{|c @{ : } c c c c c|}
        \hline
        $n$ &$ 6 $ &$ 7 $ &$ 8 $ &$ 9 $ &  $10$\\
        $c_0$ & 1.47671 & 1.47686 & 1.47694 & 1.47697 & 1.47699 \\
        $S$ &6.4$\times 10^{-10}$ & 1.8$\times 10^{-10}$ & 8.3$\times 10^{-11}$ & 7.7$\times 10^{-11}$ & 7.5$\times 10^{-11}$ \\
        \hline
        \end{tabular}
    \end{center}
\end{table}
\begin{align}\label{eqn:fitting_curve}
C_{n}(A) = c_0 + \frac{c_1}{A} + \frac{c_2}{A^2} + ... + \frac{c_n}{A^n}.
\end{align}
This amounts to taking a Taylor expansion of $C(A)$ in $A^{-1}$. We are only interested in extracting the value that $C(A)$ tends towards for large $A$, parametrised by $c_0$. The higher order terms in $C_n(A)$ serve to separate the short-range effects from the long-range effects, and are themselves discarded. This allows for greater accuracy in determining $c_0$, which should depend only on long-range effects.\par
\begin{figure}[t]
    \vspace{-10pt}
    \centering
    \includegraphics[width=0.42\textwidth]{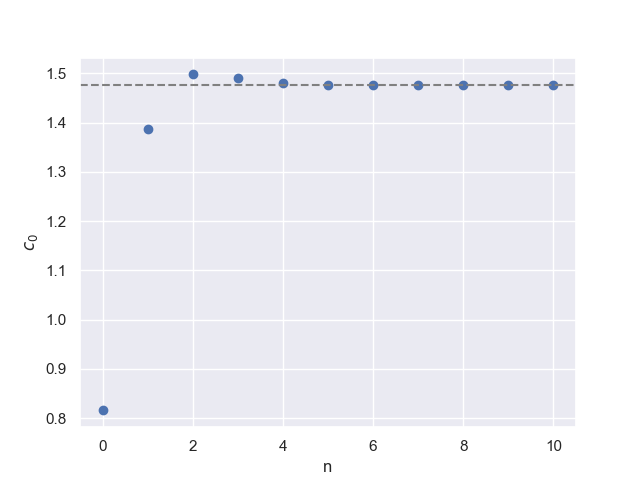}
    \captionof{figure}{Value of $c_0$ computed for a set of $C_n(A)$ fittings for $n$ between 0 and 10. The dashed line indicates the predicted value ($C = 1.4771...$).} \label{fig:fitting_plot_kak_manton}
    \vspace{+10pt}
\end{figure}

Fittings were calculated for values of $n$ between 0 and 10, and the values of $c_0$ obtained are shown in figure \ref{fig:fitting_plot_kak_manton}. This was done using the \texttt{curve\textunderscore fit} function from the \texttt{scipy.optimize} library. As $n$ is increased, and more terms are added to the fitting, $c_0$ tends towards a stable value. Table \ref{table:kink_antikink_different_fittings} shows the value of $c_0$ obtained for each $n$ tested. Furthermore, in order to assess the quality of fit, the normalised sum of squared residuals is evaluated, given by:
\begin{align}
    S =\frac{1}{N} \sum_i \left [C(A_i) - C_n(A_i) \right ]^2.
\end{align}
Here $N$ is the total number of points calculated, $C(A_i)$ is the measured value for the force coefficient at point $A_i$, and $C_n(A_i)$ is the predicted value of $C$ from the fit. This is also shown in table \ref{table:kink_antikink_different_fittings}.\par
From this information we conclude that the accuracy of the fit, as well as the validity of the model, improves for higher values of $n$. Thus, here and in the subsequent sections, we will use $C_{10}$ as the model for fitting the data, it is possible to go to higher orders, however the effect on $c_0$ is minimal and the calculations become computationally expensive.\par
To further support the validity of this choice of model, we notice that at large $n$ the higher order contributions to $C_n$ are too small to have any effect beyond extremely short distances, as shown in table \ref{table:kink_antikink_manton_example}. Even at even reasonably small distances ($A\sim 10$) the higher order contributions are negligible when compared to that from $c_0$, and so have no effect on the fitting of long range effects beyond removing the influence of the short range behaviour. The final result for $c_0$ obtained matches the predicted value of 1.4771 to a precision of four significant figures.
\begin{table}[ht]
    \begin{center}
    \caption{Fitting values for $C_{10}(A)$}\label{table:kink_antikink_manton_example}
        \begin{tabular}{|c|c|c|c|c|c|c|c|c|c|c|}
        \hline
        $c_ 0 $ &$c_ 1 $ &$c_ 2 $ &$c_ 3 $ &$c_ 4 $ &$c_ 5 $ &$c_ 6 $ &$c_ 7 $ &$c_ 8 $ &$c_ 9 $ &  $c_{10}$\\
        1.47699 & -3.1&  -2.9 & 15 & 21 & -22 & 620 & -950 & 790 &  7600 & 7900\\
        \hline 
        \end{tabular}
    \end{center}
\end{table}

\subsection{Kinks and Mirror Kinks}\label{sec:kink-mirrorkink-manton}
We now repeat the same analysis, examining the interaction between a kink and a mirror kink. The mirror kink has the same profile as a kink, but reflected over both the $x$ and $\phi$-axis. Thus we expect the correct solution to have symmetry under simultaneous $x$ and $\phi$-reflection and to pass through the origin. We propose a solution constructed by gluing together the profile of an accelerating kink to that of an accelerating mirror kink.\par
Now we return to eqn. \ref{force_total}. The time derivative and potential term are both zero, so the total force acting on the kink will be
\begin{align}\label{force_mirror_kink}
    F =\left . \frac{1}{2}\left ( \frac{\partial\phi}{\partial x}\right )^2 \right |_{x=0}.
\end{align}
We make the same substitution as in the antikink case, $\phi(x,t) = \chi\left( x - A(t)\right)$, however this time we choose positive $\ddot{A} = a$ to get
\begin{equation}
    \chi'' + a\chi' - \frac{\drv V }{\drv \chi}=0.
\end{equation}
Under the mechanical reinterpretation, this corresponds to a particle moving under a positive drag term in the inverted potential shown in fig. \ref{mech}. In this case, a valid kink solution will correspond to the particle starting at point $\chi = 0$ with positive velocity $\chi'(-A)$. The particle then loses just enough energy over its motion to asymptotically approach the maximum at $\chi = 1$.\par
As before, we assume that the drag term acts primarily over the long-range tail of the kink, making the approximation $\chi ' = \frac{dW}{d\chi}$ and integrating to get
\begin{align}
    \chi' = \sqrt{\chi^4 + \frac{4a}{15}}.
\end{align}
We expect that this solution has $\chi(-A) = 0$ and that  diverges at $\chi(0)$, thus we arrive at the integral
\begin{align}
    \int^{\infty}_{0}\frac{\drv \chi}{\sqrt{\chi^4 + \frac{4a}{15}}} = A.
\end{align}
Substituting $\chi =\left( \frac{4a}{15}\right)^{\frac{1}{4}}\lambda$ gives
\begin{equation}
    \int_{0}^{\infty}\frac{\drv \lambda}{\sqrt{\lambda^4 + 1}} = \left( \frac{4a}{15}\right)^{\frac{1}{4}} A.
\end{equation}
This is another elliptic integral of the first kind, and can be evaluated to give \cite{gradshtein_table_2007}
\begin{align}
    \left( \frac{4a}{15}\right)^{\frac{1}{4}} A = \frac{\Gamma\left ( \frac{1}{4}\right ) ^2}{4 \sqrt{\pi}}.
\end{align}
Finally, we plug this result into eqn.~\ref{force_acceleration} to get
\begin{equation} \label{prediction_mirror}
    F =\frac{1}{2A^4}\left (\frac{\Gamma(\frac{1}{4})^2}{4\sqrt{\pi}}  \right )^4 \simeq \frac{5.90852...}{A^4}.
\end{equation}

\subsubsection{Numerical Investigation}
\begin{figure}[t]
    \vspace{-10pt}
    \centering
    \includegraphics[width=0.42\textwidth]{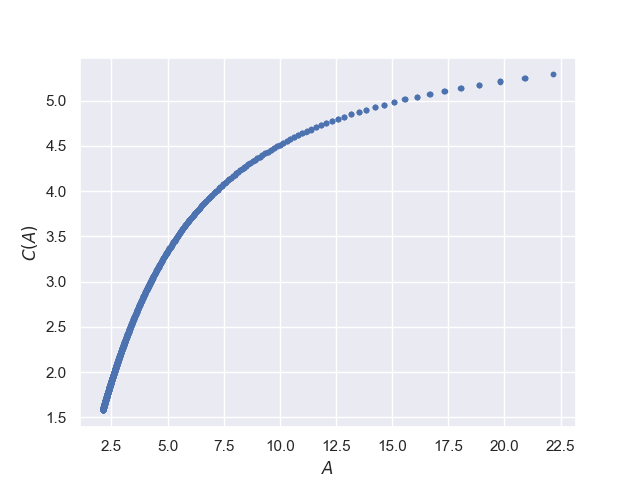}
    \captionof{figure}{Plot showing the relationship between force coefficient $C(A)$ and inter-kink separation $A$ for an interacting kink-mirror kink pair.} \label{coefficient_scatter_kmk}
    \vspace{+10pt}
\end{figure}

The numerical method here mirrors that undertaken in the case of kinks and antikinks. We start with the equation
\begin{equation}
\chi'' + a\chi' - \frac{\drv V }{\drv \chi}=0.
\end{equation}
This time, rather than picking a non-zero initial value of $\chi(-A)$, we set $\chi(-A) = 0$ and pick a starting value of $\chi'(-A)$. Then we use an identical method as before to converge on the correct value of $a$ such that we get a valid kink solution. Once we have our values for $\chi'(-A)$ and $a$, we may calculate the force acting on the kink using eqn.~\ref{force_mirror_kink}, and the distance of the kink from the midpoint between the kink and the mirror kink by looking for the point in space where the kink crosses $\chi(0) = \frac{1}{\sqrt{2}}$.\par
This was done for an array of 8000 values of $\chi'(-A)$, evenly spaced between 0.005 and 0.4. The force coefficient (eqn. \ref{force_coeff}) was calculated and the results were fitted to a curve of the form
\begin{align}
C_{\textup{10}}(A) = d_0 + \frac{d_1}{A} + \frac{d_2}{A^2} + ... + \frac{d_{10}}{A^{10}}.
\end{align}
The procedure for determining the choice of fit closely mirrors that presented in \textsection \ref{sec:numerical_manton_kak}, and so we do not restate it here. The analysis gave a $d_0$ value of 5.9031, which matches the predicted value of 5.9085 to two significant figures.\par

\section{Forces from Perturbed Equation of Motion}\label{sec:gonzales}
We now describe a different method for calculating the force between an interacting pair of kinks, following from work in \cite{gonzalez_solitary_1987,gonzalez_kinks_1989} where the $\frac{1}{A^4}$ dependence was determined but the coefficient was not. We start with the potential
\begin{align}
    V = \frac{1}{2} \left ( 1 - \phi^2\right)^2 \phi^4.
\end{align}
It has been shown that this potential is unable to support static solutions containing more than one kink. This is because fields containing multiple kinks will always experience acceleration due to interactions between the kinks \cite{manton_forces_2019}.\par
\begin{figure}[t]
    \vspace{-10pt}
    \centering
    \includegraphics[width=0.42 \columnwidth]{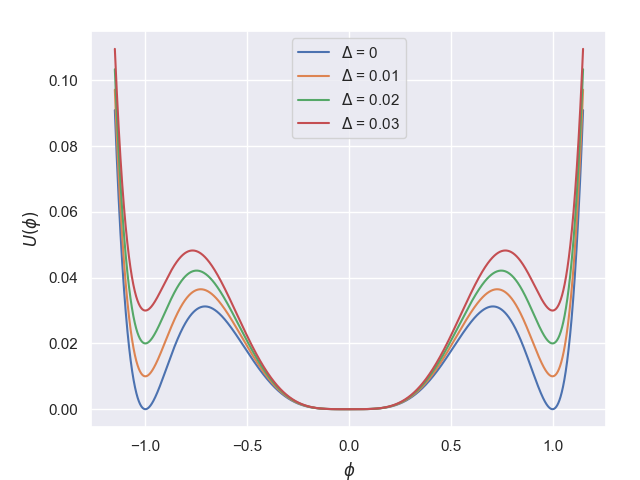}
    \caption{The potential $U(\phi)$ in eqn.~\ref{eqn:potential_delta}, with $\Delta = 0,\, 0.01,\,0.02$ and $0.03$.} 
    \label{potential_delta}
    \vspace{+10pt}
\end{figure}

To study the interacting kinks, we now modify the potential. We add a small extra term that preserves the fourth order zero at the origin and displaces the value of the potential at the minima at $\phi = \pm 1$ by a small amount $\Delta$. This gives us a new potential,
\begin{align}\label{eqn:potential_delta}
    U(\phi) = \frac{1}{2}\phi^4 \left (1-\phi^2\right )^2 - \Delta\phi^4 \left (  2 \phi^2 - 3 \right),
\end{align} 
which is plotted in fig. \ref{potential_delta}.\par
This change has the effect of exerting a force on any kinks in the system. If $\Delta$ is positive, then kinks are accelerated towards the side that is in the vacuum $\phi = \pm1$. Thus, kinks are accelerated to the right, antikinks and mirror kinks to the left, locally enlarging the region where $\phi$ is close to zero. The strength of the force exerted on a kink is equal to the value of $\Delta$. This may be intuitively understood in terms of the energy density of the vacuum on one side of the kink, which also equals $\Delta$. Sliding a kink towards the right will remove a region of energy density $\Delta$ and create an equal-sized region of energy density zero. Thus the force experienced by the kink is $\Delta$. \par
In this regime, solitary kinks are no longer static solutions of the field equations. Instead we have static solutions containing bound pairs of kinks. We look separately at the case of kinks with antikinks, and kinks with mirror kinks.

\subsection{Kinks and Antikinks}
If $\Delta >0$ the stationary state of the theory will contain an interacting kink-antikink pair. This is because the kinks and antikinks attract one another. As long as the kink is on the right and the antikink is on the left there will also be the force -- due to $\Delta$ -- that pushes them apart. The static solution corresponds to the case where the kinks are at the right distance to ensure that the force pushing them apart is balanced against the force pulling them together. Note that this configuration is unstable. If the kinks are not at precisely the correct distance from one another they will be accelerated away from the static separation distance. \par
\begin{figure}[t]
    \centering
    \includegraphics[width=0.4 \columnwidth]{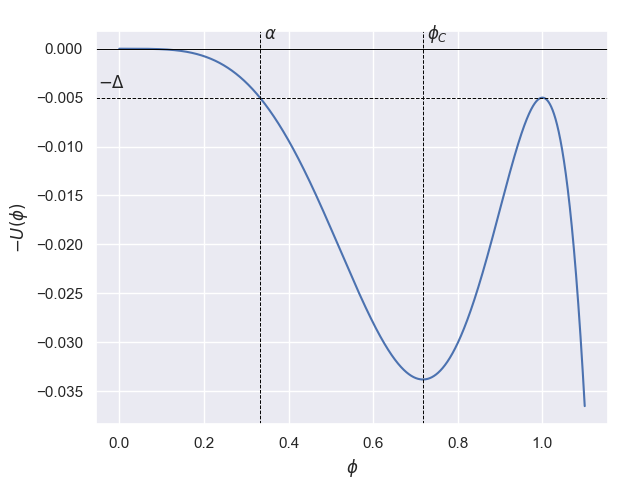}
    \caption{Plot of the potential felt by the particle in the mechanical interpretation of eqn.~\ref{eqn:mech_EOM}, with $\Delta$ = 0.005. $\alpha$ marks the smallest value of $\phi$ reached by the particle over its motion. $\phi_C$ marks the minimum of the potential, this is the value of $\phi$ that we consider to be at the centre of the kink.} 
    \label{fig:inverted_potential}
\end{figure}

To derive the profile of the bound kink-antikink pair, we examine the static equation of motion,
\begin{align}\label{eqn:mech_EOM}
    \frac{\drv^2 \phi}{\drv x^2} - \frac{\drv U}{\drv \phi} = 0.
\end{align}  
In the context of the mechanical interpretation, this can be understood as the equation of motion for a particle moving in the inverted potential $-U$, shown in fig.~\ref{fig:inverted_potential}.\par
\begin{figure}[t]
    \centering
    \includegraphics[width=0.4\columnwidth]{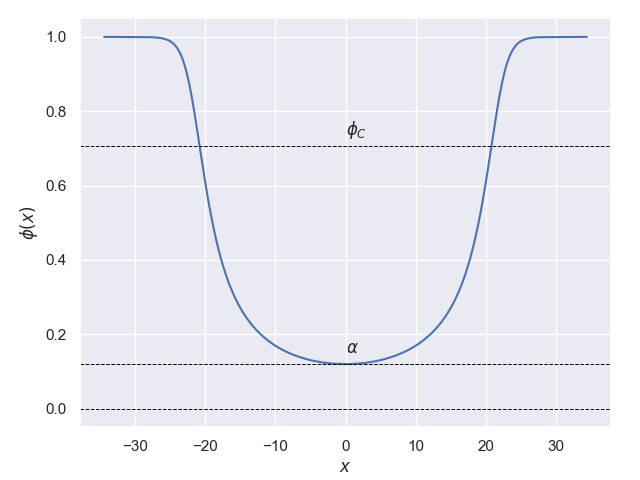}
    \caption{A bound kink-antikink pair for $\Delta$ = 0.0001. The value $\phi_C$ marks the point where the kinks cross the minimum of the inverted potential energy $-U(\phi)$, which we consider to be the centre of a kink. The minimum value of $\phi$ is marked as $\alpha$. The inter-kink half-separation can be read off the graph as $\approx 20$.} 
    \label{fig:kink_antikink_pair}
\end{figure}

The solution corresponding to a bound kink and antikink is equivalent to the particle starting at `position' $\phi = 1$ with infinitesimal velocity in the negative $\phi$ direction. The particle then almost makes it to the maximum of the potential at $\phi = 0$, but does not quite reach it, and so has its direction reversed at $\phi = \alpha$ after which it returns to its starting position at $\phi = 1$. This trajectory, and thus the shape of the bound kink-antikink pair is shown in fig.~\ref{fig:kink_antikink_pair}. We consider the position of each kink to be the point at which the field crosses the maximum of the potential energy, which we will label $\phi_C$. For a static kink this is at $\phi = \frac{1}{\sqrt{2}}$ (eqn. \ref{eqn:kink_center}), however the value is displaced slightly in the case of the modified potential. In the following work we will label this point as $\phi_C$.\par

We now wish to find an expression for the position of the kink, $A$,  as a function of $\Delta$. This can be done by integrating the equation of motion (\ref{eqn:mech_EOM}) between the points $(x = 0,\, \phi = \alpha,\, \phi' = 0)$ and $(x = A ,\, \phi = \phi_C,\, \phi' >0 )$. We integrate
\begin{align}\label{eqn:int_eom}
    \frac{\drv^2 \phi}{\drv x^2} \frac{\drv \phi}{\drv x} = \frac{\drv U}{\drv x }
\end{align}
to get
\begin{align}
    \frac{1}{2} \left(\frac{\drv \phi(x)}{\drv x}\right)^2 = U\left [ \phi(x) \right ]- U\left [ \alpha\right ],
\end{align}
which is then further integrated and rearranged to give
\begin{align}\label{eqn:main_integral}
    A = \frac{1}{\sqrt{2}}\int_{a}^{\phi_C}\frac{\drv \phi}{\sqrt{U\left( \phi \right ) -  U\left( \alpha \right ) }}.
\end{align}
Since $U\left( \alpha \right ) = \Delta$, we may re-express the term in the square root as
\begin{align}
    U\left( \phi \right ) -  U\left( \alpha \right ) = \frac{1}{2}\phi^4 \left (1-\phi^2 \right )^2 - \Delta \left (  2 \phi^6 - 3\phi^4\right) - \Delta.
\end{align}
We now factorise this polynomial to rewrite the integral as
\begin{align}\label{eqn:integral_kink_antikink}
    A =  \int_{\alpha}^{\phi_C} \frac{\drv \phi}{ (1-\phi^2) \sqrt{\phi^4 - 4\Delta \phi^2 -2\Delta } }.
\end{align}
Before we continue we must derive expressions for $\alpha$ and $\phi_C$. For $\alpha$, we find the zeros of $U(\phi) - \Delta$. This can be rewritten as
\begin{align}
    \left (1-\phi^2 \right )^2 \left (\phi^4  - 4\Delta \phi^2 -2\Delta \right ) = 0.
\end{align}
This has two obvious solutions at $\phi = \pm1$. The second polynomial has roots $\phi^2 = \alpha^2$ and $\phi^2 = -\beta^2$, with
\begin{align}
    \alpha^2 &= 2\Delta + \sqrt{2\Delta(1+2\Delta)}\\
    \beta^2 &= -2\Delta + \sqrt{2\Delta(1+2\Delta)}
\end{align}
For $\phi_C$ we solve
\begin{align}
    \frac{\drv U }{\drv \phi} = 0,
\end{align}
to get the solution
\begin{align}\label{eqn:phi_C}
    \phi_C = \frac{1}{2}\sqrt{ 2 + 12\Delta}.
\end{align}
Thus we end up with the following integral
\begin{align}
    A =  \int_{\alpha}^{\phi_C} \frac{\drv \phi}{ (1-\phi^2) \sqrt{(\phi^2 - \alpha^2)(\phi^2 + \beta^2)} }.
\end{align}
We evaluate this integral in the limit of small $\Delta$. Looking at the expressions for $\alpha$ and $\beta$, we see that in this limit
\begin{equation}
    \alpha^2 \approx \beta^2 \approx \sqrt{2\Delta}
\end{equation}
and so we can re-express the integral as
\begin{align}
    A = \int_{(2\Delta)^{1/4}}^{\phi_C} \frac{\drv \phi}{ (1-\phi^2) \sqrt{\phi^4 - 2\Delta} }.
\end{align}
This integral diverges for $\Delta \rightarrow 0$; clearly the divergence occurs at the lower limit. Thus we separate the integral into two parts
\begin{align}
    A =  \int_{(2\Delta)^{1/4}}^{s} \frac{\drv \phi}{ (1-\phi^2) \sqrt{\phi^4 -2\Delta} } +  \int_{s}^{\phi_C} \frac{\drv \phi}{ (1-\phi^2) \sqrt{\phi^4 - 2\Delta} },
\end{align}
where $s$ is some small parameter we have chosen such that $(2\Delta)^{1/4} \ll s \ll 1$. The first integral here, from $(2\Delta)^{1/4}$ to $s$ is divergent as $\Delta \rightarrow 0$, but the second part is not and so may be discarded. Furthermore, now that $(2\Delta)^{1/4} \ll 1 $ and $s\ll 1$ we can assume that $(1-\phi^2) \approx 1$ in the first integral. We arrive at the following
\begin{align}
    A \sim  \int_{(2\Delta)^{1/4}}^{s} \frac{\drv \phi}{ \sqrt{\phi^4 - 2\Delta} } .
\end{align}
Now we make a substitution $\phi = (2\Delta)^{1/4} \psi$ to get
\begin{align}
    A \sim \frac{1}{(2\Delta)^{1/4}} \int_{1}^{s (2\Delta)^{-1/4}} \frac{\drv \psi}{ \sqrt{\psi^4 - 1} } .
\end{align}
The upper limit approaches infinity for small $\Delta$ and so we are left with a complete elliptic integral of the first kind. This is evaluated to give
\begin{align}
A = \frac{1}{(2\Delta)^{1/4}} \frac{1}{\sqrt{2}}\frac{\Gamma (\frac{1}{4})^2}{4\sqrt{\pi}}
\end{align}
and rearranged to arrive at the final expression
\begin{align}
\Delta = \frac{1}{8A^4} \left ( \frac{\Gamma (\frac{1}{4})^2}{4\sqrt{\pi}}\right )^4 = \frac{1.4771306...}{A^4}  .
\end{align}
Remembering that $\Delta$ is just the force of attraction between the kinks, we can see that we have recovered the identical force as in eqn. \ref{prediction}.

\subsubsection{Numerical Investigation}
To verify that the approximations made are sound, we numerically calculate the integral in eqn. \ref{eqn:integral_kink_antikink}. Values were found using the \texttt{quad} function from Python's \texttt{scipy.integrate} library. An array of 1,000,000 values of $\Delta$ were used, evenly spaced between $\Delta = 5\times 10^{-8}$ and $\Delta = 1\times 10^{-3}$.\par
As before, the values obtained for distance and force were used to calculate the force coefficient $C(A) = FA^4$ and this was fitted to a function of the form
\begin{align}\label{eqn:c10}
    C_{10}(A) = c_0 + \frac{c_1}{A} + \frac{c_2}{A^2} + ... + \frac{c_{10}}{A^{10}}.
\end{align}
We use the same procedure for choosing the fit as in \textsection \ref{sec:numerical_manton_kak}. The value of $c_0$ obtained using this method was 1.4771336, agreeing with the predicted value, 1.4771306, to six significant figures, suggesting that the approximations used were valid.

\subsection{Kinks and Mirror Kinks}\label{sec:kinks_mirror_kinks_gonzales}
\begin{figure}
   \centering
   \includegraphics[width=0.4\columnwidth]{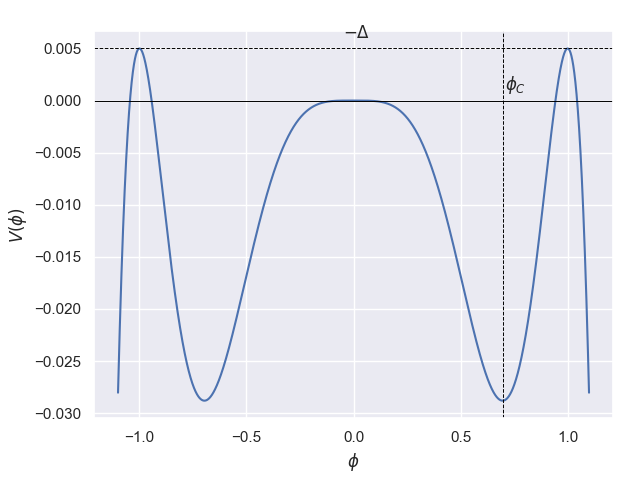}
   \caption{Plot of the potential felt by the particle in the case $\Delta = -0.005$. Once again $\phi_C$ marks the centre of the kinks.} 
   \label{mechanical_2}
\end{figure}

We can now repeat the same method to find the force of repulsion between a kink and mirror kink. We use the same modified potential, eqn.~\ref{eqn:potential_delta}, however this time $\Delta$ must be negative. This is because our kink and mirror-kink repel one another, so we must add a force of attraction between them, arriving at the mechanical potential shown in fig.~\ref{mechanical_2}.\par
\begin{figure}
   \centering
   \includegraphics[width=0.4\columnwidth]{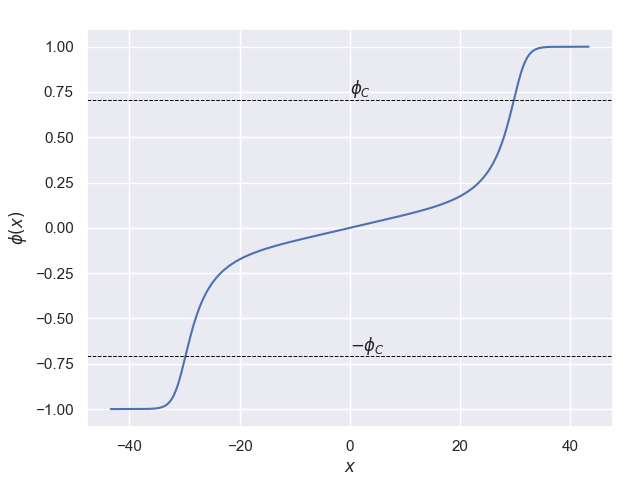}
   \caption{Plot of a bound kink-mirror-kink pair, for $\Delta = -0.0001$. The points $\phi_C$ and $-\phi_C$ mark the positions of the centre of each kink and the inter-kink half-separation can be read off as  $\approx 30$.} 
   \label{kink_mirrorkink_pair}
\end{figure}

In this case, the kink solution corresponds to the particle starting at $\phi = -1$ with infinitesimal positive velocity. The particle then passes through the point $\phi = 0$ with `velocity' $\phi'(0)=\sqrt{2\Delta}$ (obtained from conservation of kinetic and potential energy). It then crosses over $\phi_C$ and comes to rest at $\phi=1$. The resulting kink shape is shown in fig.~\ref{kink_mirrorkink_pair}.\par
Again, we must find an expression for the inter-kink distance, which is done by integrating between the points $(x=0,\, \phi=0 ,\, \phi' = \sqrt{2\Delta})$ and $(x=d/2,\, \phi=\phi_C ,\, \phi' >0)$. Expression~\ref{eqn:int_eom} integrates to give
\begin{align}
\frac{1}{2} \left(\frac{\drv \phi(x)}{\drv x}\right)^2 - \frac{1}{2} \left(\sqrt{2\Delta} \right)^2  = U\left [ \phi(x) \right ].
\end{align}
We rearrange and integrate again to get:
\begin{align}\label{eqn:kink_mirrorkink_integral}
A = \frac{1}{\sqrt{2}} \int_{0}^{\phi_C}\frac{\drv \phi}{\sqrt{U(\phi) + |\Delta|}}.
\end{align}
The term in the square root is now
\begin{align}
    \frac{1}{2} \left (1-\phi^2 \right )^2 \left (\phi^4 + 4 |\Delta| \phi^2+2 |\Delta| \right )
\end{align}
so we obtain an expression of the form
\begin{align}
A = \int_{0}^{\phi_C}\frac{\drv \phi}{\left (1-\phi^2 \right )\sqrt{\phi^4 + 4 |\Delta| \phi^2+2 |\Delta|}}.
\end{align}
We factorise the polynomial in the square root to get 
\begin{align}\label{eqn:integral_mirror_ab}
A = \int_{0}^{\phi_C}\frac{\drv \phi}{\left (1-\phi^2 \right )\sqrt{(\phi^2 - \alpha^2)(\phi^2 - \beta^2)}},
\end{align}
with
\begin{align}
    \alpha^2 &= -2|\Delta| + \sqrt{4|\Delta|^2 - 2|\Delta |}\\
    \beta^2 &= -2|\Delta| - \sqrt{4|\Delta|^2 - 2|\Delta |}.
\end{align}
$\phi_C$ is again determined by eqn.~\ref{eqn:phi_C}. Taking $\Delta$ to be small, we can set $\alpha^2 \approx -\beta^2 \approx i \sqrt{2|\Delta|}$. Thus from eqn.~\ref{eqn:integral_mirror_ab} we get
\begin{align}
A = \int_{0}^{\phi_C}\frac{\drv \phi}{\left (1-\phi^2 \right )\sqrt{\phi^4 + 2|\Delta|}}.
\end{align}
This integral diverges at the lower limit, as $\Delta \rightarrow 0 $, so we split it into two integrals.
\begin{align}
A = \int_{0}^{s}\frac{\drv \phi}{\left (1-\phi^2 \right )\sqrt{\phi^4 + 2|\Delta|}} + \int_{s}^{\phi_c}\frac{\drv \phi}{\left (1-\phi^2 \right )\sqrt{\phi^4 + 2|\Delta|}}
\end{align}
with $(2\Delta)^{1/4} \ll s \ll 1$. The second integral is not divergent so we may discard it. Furthermore, the value of $s$ is small so we may ignore the $(1-\phi^2)$ term in the first integral. Thus, after making a substitution $\phi = (2\Delta)^{1/4}\psi$ we get another complete integral of the first kind, since the upper limit approaches infinity for small $\Delta$.
\begin{align}
    A \sim \frac{1}{(2\Delta)^{1/4}} \int_{0}^{s(2\Delta)^{-1/4}} \frac{\drv \psi}{ \sqrt{\psi^4 + 1} } .
\end{align}
This integral can be evaluated and rearranged to give
\begin{align}\label{eqn:prediction}
    \Delta = \frac{1}{2 A^4} \left ( \frac{\Gamma\left ( \frac{1}{4}\right )  ^2}{4\sqrt{\pi}}\right )^4 = \frac{5.9085225...}{A^4}.
\end{align}
Again, we have arrived at an identical force as in eqn. \ref{prediction_mirror}.

\subsubsection{Numerical Investigation}
As before, we verify these results by integrating equation~\ref{eqn:kink_mirrorkink_integral} numerically. It is done using the same methods as the previous section, using 1,000,000 values of $\Delta$, equally spaced from $-0.00000005$ to $-0.001$. After finding the force coefficient, $C(A)$, and applying the same fitting as eqn.~\ref{eqn:c10} we obtain a value of $d_0 = 5.908496$. These results match the prediction (eqn. \ref{eqn:prediction}) to five significant figures, confirming that the method outlined in \textsection\ref{sec:kinks_mirror_kinks_gonzales} is sound.

\section{Conclusions and Outlook}

We have calculated the forces between the kinks that emerge in a special $\phi^8$ theory, where the kinks have large spatial extent and are able to interact over long distances. This was done using two different methods. The first, following \cite{manton_forces_2019}, looked at approximately solving the equation of motion for an accelerating kink. Collective coordinate methods were used to reduce the equation of motion to a modified Bogomolny equation that could be solved approximately. The second method adapted the techniques in \cite{gonzalez_kinks_1989}. A small perturbation was added to the potential in order to construct static solutions containing pairs of interacting kinks. In both cases, the strength of force acting between kink-antikink and kink-mirror kink pairs at large separation was found to have a $\frac{1}{A^4}$ dependence on the inter-kink half-separation and the prefactor was calculated. Furthermore we demonstrated analytically that both methods make identical predictions for the prefactor.\par
In both cases, we supported our calculations with numerics, calculating the strength of the force between the kinks. In the first section this was done by numerically solving the modified Bogomolny equation and extracting the separation and force between the kinks from the solution. This required us to tune the input parameters of the equation to obtain a valid kink solution, a step that substantially increased the computing time. In the second case we were able to reduce the problem to an integral that could be numerically evaluated. The second method proved much less computationally expensive, allowing us to sample the parameter space extremely finely. This meant that the results obtained using the second method had a greater accuracy, matching the prediction to six and five significant figures for the kink-antikink and kink-mirror kink cases respectively. In comparison the results from the first method matched the prediction to three and two significant figures for the kink-antikink and kink-mirror kink cases. These results are shown in table \ref{table:results}. This suggests that the second method is preferable when calculating the long-range force between pair of kinks. \par
\begin{table}[ht]
    \begin{center}
    \caption{Predictions and numerical results for the force coefficient.  }\label{table:results}
        \begin{tabular}{|c|c|c|c|}
        \hline
         &Predicted Value & First Method & Second Method  \\
        Kink - Antikink & 1.477131 &  1.47699 & 1.477133  \\
        Kinks - Mirror Kink &5.90852&  5.903 & 5.80849 \\
        \hline 
        \end{tabular}
    \end{center}
\end{table}
Potential further work could be to generalise the methods presented here to different systems, in which the potentials could have higher-order minima. We expect that the techniques presented will be transferable to a number of variations on the theory. Our work is limited to cases where the kinks are well-separated, thus it would be interesting to investigate the interactions between kinks when the condition of large separation has been relaxed. Additionally we have only calculated the force acting between initially static field configurations, which represent an extremely limited range of the configurations available to such a system. It would be valuable to investigate how these forces change for kinks in motion relative to one another. The interactions between kinks in our theory with radiation would be worth studying, since there is evidence that such interactions can display many interesting characteristics \cite{forgacs_negative_2008,romanczukiewicz_could_2017}.

\section{Acknowledgments}

The work presented here develops on an essay submitted for the completion of Part III of the Mathematical Tripos at Cambridge University under the supervision of Professor Nick Manton. The author would like to thank Professor Manton for setting the interesting topic, and for his generous and meticulous feedback on the work. Furthermore, the author would like to thank Dr. Derek Lee of Imperial College London for his helpful advice regarding some of the calculations presented above.

\bibliography{ref_list.bib}

\end{document}